\documentclass{amsart}
\date{\today}
\usepackage{amssymb}
\usepackage{latexsym}
\newcommand{\Z}{{\mathbb Z}}
\newcommand{\R}{{\mathbb R}}

\let\Re=\undefined\DeclareMathOperator*{\Re}{Re}

\newtheorem{theorem}{Theorem}

\newtheorem{lemma}{Lemma}[section]
\newtheorem{prop}[lemma]{Proposition}
\newtheorem{coro}{Corollary}

\newcounter{smalllist}
 \makeatletter\renewcommand{\thesmalllist}{\@alph\c@smalllist}\makeatother

\begin{document}
\title{Schr\"odinger Operators With Few Bound States}

\author{David Damanik}

\address{Mathematics 253--37, California Institute of Technology, Pasadena, CA 91125, USA}
\email{damanik@caltech.edu}

\author{Rowan Killip}

\address{Department of Mathematics, University of California, Los Angeles, CA 90055, USA}
\email{killip@math.ucla.edu}

\author{Barry Simon}

\address{Mathematics 253--37, California Institute of Technology, Pasadena, CA 91125, USA}
\email{bsimon@caltech.edu}

\thanks{D.\ D.\ was supported in part by NSF grant DMS--0227289.}
\thanks{R.\ K.\ was supported in part by NSF grant DMS--0401277.}
\thanks{B.\ S.\ was supported in part by NSF grant DMS--0140592.}

\begin{abstract}
We show that whole-line Schr\"odinger operators with finitely many bound states have no
embedded singular spectrum. In contradistinction, we show that embedded singular spectrum
is possible even when the bound states approach the essential spectrum exponentially
fast.

We also prove the following result for one- and two-dimensional Schr\"odinger operators,
$H$, with bounded positive ground states: Given a potential $V$, if both $H\pm V$ are
bounded from below by the ground-state energy of $H$, then $V\equiv 0$.
\end{abstract}

\maketitle

\section{Introduction}

This paper has its roots in the following result of Killip and Simon \cite{ks}: A
discrete whole-line Schr\"odinger operator has spectrum $[-2,2]$ if and only if the
potential vanishes identically.

To be more precise, given a potential $V: \Z \to \R$, we define the Schr\"odinger
operator
\begin{equation}\label{E:discoper}
[h_V \phi](n) = \phi(n+1) + \phi(n-1) + V(n) \phi(n)
\end{equation}
on $\ell^2(\Z)$. The theorem mentioned above,
\cite[Theorem~8]{ks}, says that $\sigma(h_V) \subseteq [-2,2]$
implies $V \equiv 0$. A simple variational proof of this theorem
was given in \cite{dhks}, where the result was also extended to
two dimensions; it does not hold in three or more dimensions, nor
on the half-line.

It was also shown in \cite{dhks} that, for bounded potentials, the essential spectrum of
$h_V$ is contained in $[-2,2]$ if and only if $V \to 0$. This shows that $\sigma_{{\rm
ess}}(h_V) = [-2,2]$ in this case.

Damanik and Killip, \cite{dk}, investigated half-line Schr\"odinger operators with
spectrum contained in $[-2,2]$. By a half-line Schr\"odinger operator we mean an operator
of the form \eqref{E:discoper}, acting on $\ell^2(\Z^+)$, with $\phi(-1) = 0$. We will
denote this operator by $h_V^+$. It was shown that if $\sigma(h_V^+) = [-2,2]$, then
$h_V^+$ has purely absolutely continuous spectrum. Using the methods developed to prove
this result, it was also shown that half-line Schr\"odinger operators with finitely many
bound states (i.e., eigenvalues lying outside $[-2,2]$) have purely absolutely continuous
spectrum on $[-2,2]$. See also Simon \cite{s2}.

Several people have asked, in private, whether one can deduce absence of embedded
singular spectrum from
$$
\sum_n \sqrt{|E_n| - 2} < \infty.
$$
(This condition often arises as a natural borderline condition for regular behavior of
the spectral measure near the points $-2$ and $2$; see for example \cite{py,sz}.  It is
equivalent to the convergence of the Blaschke product associated with the eigenvalues
after the cut plane is mapped to the unit disk by the inverse of $z \mapsto z+z^{-1}$.)

In fact, embedded singular spectrum can occur even when the bound states approach $\pm 2$
exponentially fast. This is our first result.

\begin{theorem}\label{T:embev}
There is a discrete half-line Schr\"odinger operator with zero as
an eigenvalue, whose bound states obey $|E_n| - 2 \le c^n$ for
a suitable $c < 1$.
\end{theorem}

\noindent\textit{Remark.} As we will show, $|E_n| - 2$ also admits an exponential lower
bound. We do not know whether the existence of embedded singular spectrum places a lower
bound on the rate at which the eigenvalues can approach $[-2,2]$. On the other hand,
recent work of Damanik and Remling, \cite{dr}, shows that finiteness of the $p$-th
eigenvalue moment implies that the embedded singular spectrum must be supported on a set
of Hausdorff dimension $4p$.

\medskip

The Damanik--Killip paper, \cite{dk}, discussed earlier also considered half-line
continuum Schr\"odinger operators with a Dirichlet boundary condition. That is, given a
potential $V$ we define an operator on $L^2(\R^+)$ by
\begin{equation}\label{E:contop+}
[H_V^+ \phi] (x) = -\phi''(x) + V(x) \phi(x)
\end{equation}
with $\phi(0)=0$. In this case it was proved that if $V$ belongs to the space
$\ell^\infty(L^2)$ of uniformly locally square-integrable functions and both $H_{V}^+$
and $H_{-V}^+$ have only finitely many bound states, then both have purely absolutely
continuous spectrum on $[0,\infty)$. This is the continuum analogue of the
Damanik--Killip theorem discussed above, as we will now explain: the unitary map
$\phi(n)\mapsto (-1)^n\phi(n)$ conjugates $h_{-V}$ to $-h_{V}$; consequently, $h_V$ has
finitely many bound states if and only if both $2-h_{\pm V}$ have only finitely many
bound states below zero.

It is natural to ask what one may say for whole-line Schr\"odinger operators with
finitely many bound states. (We write $H_V$ for the operator on $L^2(\R)$ defined through
\eqref{E:contop+}.)  The analogous results hold:

\begin{theorem}\label{T:wlfmbs}
If the operator $h_V$ has only finitely many bound states, then it
has purely absolutely continuous spectrum of multiplicity $2$ on
$[-2,2]$.
\end{theorem}

\begin{theorem}\label{T:wlfmbscont}
Suppose $V \in \ell^\infty(L^2)$. If both $H_V$ and $H_{-V}$ have only finitely many
eigenvalues below energy zero, then both operators have purely absolutely continuous
spectrum of multiplicity $2$ on the interval $[0,\infty)$.
\end{theorem}

Recall that if the operators $h_{\pm V}$ are required to have no bound states above $2$,
then $V$ must be identically $0$---this is the result of Killip--Simon. We will prove the
continuum analogue of this result by revisiting the variational approach of \cite{dhks}.
Moreover, the refinements we introduce permit us to treat more general background
potentials.

We begin with some notation. Expanding our previous usage, we let $H_V$ denote the
Schr\"odinger operator with potential $V$ in $L^2(\R^d)$ for any dimension $d$. By a
ground state for $H_V$, we mean a (distributional) solution of
$-\nabla^2\psi+V\psi=E\psi$ where $E$ is the ground-state energy, that is, the infimum of
the spectrum.  We use the analogous terminology in the discrete case; however, here we
define the ground-state energy to be the supremum of the spectrum.

\begin{theorem}\label{T:gsdisc}
Let $d = 1$ or $2$. Suppose that $Q,V$ are functions on $\Z^d$ such that the operator
$h_Q$ has a bounded positive ground state and both $h_{Q \pm V}$ are bounded above by the
ground state energy of $h_Q$. Then $V \equiv 0$.
\end{theorem}

\begin{theorem}\label{T:gscontgen2}
Let $d = 1$ or $2$. Suppose that $Q \in L^2_\mathrm{loc}(\R^d)$ and the operator $H_Q$
has a bounded positive ground state. If $V \in L^2_\mathrm{loc}(\R^d)$ and both $H_{Q \pm
V}$ are bounded below by the ground state energy of $H_Q$, then $V \equiv 0$.
\end{theorem}

As noted already in \cite{dhks}, these theorems fail for $d \ge 3$ even when $Q \equiv
0$. This is an immediate consequence of the Cwikel--Lieb--Rosenblum inequality. However,
the presence of absolutely continuous spectrum may still be deduced in some situations as
a recent result of Safronov shows \cite{saf}.

The existence of a positive ground state holds under fairly general conditions; see Simon
\cite{s3}. Moreover, it is not hard to see that, for periodic potentials $Q$, the ground
state energy corresponds to zero quasi-momentum (this can also be derived directly from
more general results of Agmon \cite{a}) and hence the operator $H_Q$ has a bounded
positive ground state. Thus, as an application of Theorem~\ref{T:gscontgen2}, we may
deduce

\begin{coro}\label{T:gscont}
Let $d = 1$ or $2$ and suppose that $Q \in L^2_\mathrm{loc}(\R^d)$ is periodic. If $V \in
L^2_\mathrm{loc}(\R^d)$ and both $H_{Q \pm V}$ are bounded below by the ground state
energy of $H_Q$, then $V \equiv 0$.
\end{coro}

The particular case $d = 1$ and $Q \equiv 0$ yields the continuum analogue of the
Killip--Simon theorem mentioned at the beginning of the introduction.  The discrete
analogue of the corollary also holds and therefore extends the original Killip--Simon
result to the case of periodic background.

Y.~Pinchover has explained to us that Theorem~\ref{T:gscontgen2} represents a statement
about the criticality of $H_Q$ (cf.~\cite{p1,p2}).  He also pointed out that through
these techniques, Corollary~\ref{T:gscont} can be derived from Theorem~2 of
\cite{Pinsky}.

We will use Dirac notation for the inner product of the underlying Hilbert space
$\mathcal{H}$. In particular, if $H$ is a self-adjoint operator in $\mathcal{H}$ and
$\phi,\psi$ belong to the form domain of $H$, then we write the quadratic form associated
with $H$ as $\langle \phi | H | \psi \rangle$.

The organization is as follows. In Section~\ref{S:2} we discuss an example with an
embedded eigenvalue and exponential bound state decay and, in particular, prove
Theorem~\ref{T:embev}. Theorems~\ref{T:wlfmbs} and \ref{T:wlfmbscont} are proven in
Section~\ref{S:3}. Finally, we study perturbations of Schr\"odinger operators with
positive ground states in Section~\ref{S:4} and obtain Theorems~\ref{T:gsdisc} and
\ref{T:gscontgen2}.

\medskip

\noindent\textit{Acknowledgment.} It is our pleasure to thank Y.~Pinchover and
A.~V.~Sobolev for useful comments.

\section{Embedded Eigenvalue and Exponential Bound State Decay}\label{S:2}

The following result contains Theorem~\ref{T:embev} and also provides an exponential
lower bound.

\begin{theorem}\label{T:ex}
There is a discrete half-line Schr\"odinger operator that has zero as an eigenvalue and
its eigenvalues $\{E_n\}$ outside $[-2,2]$ obey $b^n \lesssim |E_n| - 2 \lesssim c^n$ for
suitable $b \le c < 1$.
\end{theorem}

\noindent\textit{Remark.} For functions $f,g$, we write $f \lesssim g$ if $f/g$ is
bounded.

\medskip

We will revisit an example of Wigner--von Neumann type discussed in \cite{dk}. It is
roughly of the form $V(n) \sim \lambda (-1)^n n^{-1}$. It follows from \cite{dhks} that
$\lambda$ must be of magnitude greater than one in order for the operator to have
infinitely many eigenvalues outside $[-2,2]$.

By Weyl's theorem, the spectrum outside $[-2,2]$ consists of eigenvalues, $E_n(V)$, that
can accumulate only at $\pm 2$. We choose an ordering such that $|E_1(V)| \ge |E_2(V)|
\ge \cdots$.

\begin{proof}[Proof of Theorem~\ref{T:ex}.]
Fix $\alpha>1/2$ and define $\psi:\Z^+ \to \R$ as follows: the absolute value is given by
$|\psi(n)| = (n+1)^{-\alpha}$ and the sign depends on the value of $n$ mod $4$ with the
pattern $+,+,-,-,\ldots$.  Notice that $\psi$ is square-summable and so a zero-energy
eigenfunction for the operator $h_V$ with potential given by
$$
V(n) = -\frac{\psi(n+1) + \psi(n-1)}{\psi(n)}
$$
for $n\geq1$ and $V(0)=-\psi(1)/\psi(0)$. Clearly,
\begin{equation}\label{E:form}
V(n) = -2\alpha(-1)^n n^{-1} + O(n^{-2}).
\end{equation}

We now turn to the main part of the proof: controlling the bound states of $h_V$. Both
inequalities rely on results relating operators with sign indefinite potentials to those
with sign-definite potentials.  We begin with upper bounds.

It was shown in \cite{dhs} (see, in particular, Eq.~(4.18)) that if $V^\pm$ are defined
by
\begin{equation}\label{E:vpmdef}
V^\pm (n) = \pm 2 \bigl[ F(n)^2 + F(n+1)^2 \bigr] \quad \text{with} \quad F(n) = -
\sum_{j = n}^\infty V(j),
\end{equation}
then
$$
2 - h_V \ge \tfrac12 \left( 2 - h_{V^+} \right) \quad \text{and} \quad 2 + h_V \ge
\tfrac12 \left( 2 + h_{V^-} \right).
$$
Thus it suffices to bound the eigenvalues for $h_{V^\pm}$.  To do this we employ the
Jacobi matrix analogue of the Bargmann bound (see \cite[Theorem~A.1]{hs}):
$$
\#\{|E_n(V^\pm)| \ge \lambda+2\} \lesssim \sum (n+1) \bigl||V^\pm(n)|-\lambda\bigr|_+ .
$$
where $|x|_+ = \max\{x,0\}$.  In our case, $|V^\pm(n)| \lesssim (n+1)^{-2}$ and so
$\#\{|E_n(V^\pm)| \ge \lambda+2\} \lesssim \log(\lambda)$, or equivalently, there exists
$0<c<1$ so that $|E_n(V^\pm)| - 2 \lesssim c^n$.

We now turn to a proof of the lower bound for $|E_n(V)|$. We will employ some ideas and
results of \cite{dhks}. From Eq.~(1.7) of that paper we see that in order to prove
$|E_n(V)| - 2 \gtrsim b^n$ for some $0< b < 1$, it suffices to find trial functions
$\varphi_n$, whose supports are disjoint, such that
\begin{equation}\label{E:vtildeest}
\langle \varphi_n | h_{\tilde{V}} - 2 | \varphi_n \rangle \gtrsim b^{n}.
\end{equation}
where $\tilde V (n) = \frac{1}{4} V(n)^2$.  Note that from \eqref{E:form},
$$
\tilde{V}(n) = \alpha^2 n^{-2} + O(n^{-3}).
$$
For convenience, we pick $\alpha \ge \sqrt{7}$ (the construction below can be modified to
accommodate any $\alpha > 1/2$). Let $m_n = 8^n$. The trial function $\varphi_n$ is then
the function which is $1$ at $m_n$, has constant slope on the intervals
$[\frac{m_n}{4}-1, m_n]$ and $[m_n, \frac{3m_n}{2} + 1]$, and vanishes outside the
interval $[\frac{m_n}{4}-1,\frac{3m_n}{2}+1]$. Mimicking the arguments from the proof of
\cite[Theorem~5.5]{dhks}, we see that
$$
\langle \varphi_n | h_{\tilde{V}} - 2 | \varphi_n \rangle \ge \frac{1}{m_n}
$$
and hence \eqref{E:vtildeest} holds with $b = 1/8$. This concludes the proof.
\end{proof}

Basically, the argument that the eigenvalues have geometrically fast approach to $\pm 2$
comes from the quadratic mapping \eqref{E:vpmdef} and the fact that for supercritical
$r^{-2}$ potentials, the approach is geometric. This was shown in the continuum case by
Kirsch and Simon \cite{ks2} with explicit constants. It should be possible to compute
$\lim (|E_n| - 2)^{1/n}$ in the discrete setting along similar lines. Bounds of this form
can be used in the study of the Efimov effect; see Tamura \cite{t}, for example. The fact
that there are infinitely many bound states for coupling above a critical value was shown
in the discrete case by Na\u\i man \cite{n}.

\section{Whole-Line Operators With Finitely Many Bound States}\label{S:3}

The purpose of this section is to prove Theorems~\ref{T:wlfmbs} and \ref{T:wlfmbscont}.

By restricting a whole-line operator $h_V$ to $\ell^2(\Z^\pm)$ we
obtain two half-line operators, which we denote by $h_V^\pm$.
(Here $\Z^+ = \{ 0,1,2,\ldots \}$ and $\Z^- = \{ -1,-2,-3,\ldots
\}$.)

If $h_V$ has finitely many bound states, then so do both $h_V^\pm$
because their direct sum is a finite-rank perturbation of $h_V$.
Thus, it follows from \cite{dk} that both $h_V^\pm$
have purely absolutely continuous spectrum (essentially supported)
on $[-2,2]$. Using the finite-rank perturbation property again, it
follows that $h_V$
has absolutely continuous spectrum of multiplicity two
(essentially supported) on $[-2,2]$.

Singular spectrum, or its absence, is not stable under finite-rank perturbations. We will
revisit the half-line proof from \cite{dk}, which proceeded through controlling the
behavior of solutions and then applying the Jitomirskaya--Last version, \cite{jl}, of the
Gilbert--Pearson theory of subordinacy, \cite{g,gp}. The whole-line extension of the
Jitomirskaya--Last result that we need can be found in \cite{dkl}. We begin by recalling
the necessary results from \cite{dk,dkl}.

Let us write $\psi_\theta$ for the solution of
\begin{equation}\label{E:psiE}
\psi(n+1) + \psi(n-1) + V(n) \psi(n) = E \psi(n)
\end{equation}
that obeys the initial condition
$$
\psi_\theta(-1) = \sin (\theta), \;\; \psi_\theta(0) = \cos (\theta).
$$

\begin{prop}\label{P:solest}
Suppose that the operator $h_V^+$ has only finitely many
eigenvalues outside $[-2,2]$. Then for any energy $0<|E|<2$ and any $\eta > 1/\sqrt{2}$,
\begin{equation}\label{E:onebound}
 n^{-\eta} \lesssim \bigl| \psi_\theta(n) \bigr|^2 +
\bigl|\psi_\theta(n+1) \bigr|^2 \lesssim n^\eta
\end{equation}
for all $\theta$ and $n>0$ .   The implicit constants depend on $E$ and $\eta$, but not
$\theta$. For $E=0$, \eqref{E:onebound} holds with $\eta=1$.
\end{prop}

\begin{proof}
This follows from Corollary~4.6 and Proposition~5.2 of \cite{dk}.  While the statement
given there does not describe the dependencies of the constants, they can be deduced
readily from the proof.
\end{proof}

\begin{prop}\label{P:zeroH}
Suppose that the operators $h_V^\pm$ have only finitely many eigenvalues outside $[-2,2]$.
Then the set of energies in $(-2,2)$ for which \eqref{E:psiE} has unbounded solutions is of
Hausdorff dimension zero.
\end{prop}

\begin{proof}
As an unbounded solution of \eqref{E:psiE} must be unbounded on one side of the
origin, we may apply the half-line results of \cite{dk}, specifically, Corollary~4.6
and Proposition~5.5.
\end{proof}

The following result can be found in \cite{dkl}:

\begin{prop}\label{P:dklresult}
Suppose that for some $\eta < 1$ and each energy $E$ in a bounded set $A$,
$$
n^{-\eta} \lesssim \bigl| \psi_\theta(n) \bigr|^2 +
  \bigl| \psi_\theta(n+1) \bigr|^2 \lesssim n^\eta,
$$
for all $\theta$ and $n > 0$. Then any spectral measure for $h_V$ gives no weight to
subsets of $A$ of Hausdorff dimension less than $1 - \eta$.
\end{prop}

We now complete the proof of Theorem~\ref{T:wlfmbs} by putting these
ingredients together.

\begin{proof}[Proof of Theorem~\ref{T:wlfmbs}.]
We saw above that the absolutely continuous spectrum of $h_V$ is
essentially supported on $[-2,2]$ and has multiplicity two. Also, by
Proposition~\ref{P:solest}, $0$ is not an eigenvalue.

The singular part of any spectral measure for $h_V$ gives no weight to the set of
energies for which all solutions of \eqref{E:psiE} are bounded (see, e.g.,
\cite{b,s,stolz}). Thus Proposition~\ref{P:zeroH} shows that the singular part must be
supported on a set of zero Hausdorff dimension; while Propositions~\ref{P:solest}
and~\ref{P:dklresult} together with  the previous paragraph imply that it must give no
weight to any zero-dimensional subset of $(-2,2)$.  Thus it remains only to show that
$\pm2$ are not eigenvalues. To do so, we mimic the proof of Corollary~4.6 from \cite{dk}.

Assume that $E = 2$ is an eigenvalue. Then, after possibly changing the
value of $V(0)$, we see that $h_V^+$ also has an eigenvalue at $2$, but only finitely
many bound states.  This contradicts \cite[Corollary~4.6~(e)]{dk}. The same line of
reasoning works when one assumes that $E = -2$ is an eigenvalue.
\end{proof}

\begin{proof}[Proof of Theorem~\ref{T:wlfmbscont}.]
The proof is analogous to that of Theorem~\ref{T:wlfmbs}. Let us write $\psi_\theta$ for
the solution of $-\psi''(x) + V(x) \psi(x) = E \psi(x)$ that obeys the initial condition
$\psi_\theta(0) = \sin (\theta)$, $\psi_\theta'(0) = \cos (\theta)$.

The analogues of the three propositions above can be found in the literature: In place of
Proposition~\ref{P:solest} we can use Corollary~6.5(a) and Propositions~7.4 from
\cite{dk}. Similarly, Corollary~6.5(a) and Proposition~7.7 from \cite{dk} substitute
Proposition~\ref{P:zeroH}. That the continuum analogue of Proposition~\ref{P:dklresult}
also holds, was noted already in \cite{dkl}.

Finally, showing that zero is not an eigenvalue, can be effected by mimicking the proof
of \cite[Corollary~6.5(b)]{dk}.
\end{proof}

\section{Perturbations of Operators With Positive Ground States}\label{S:4}

In this section we extend the variational technique introduced in \cite{dhks}: we are
able to treat continuum Schr\"odinger operators and also allow more general unperturbed
operators, specifically, those with positive ground states. As there is greater novelty
in the continuum case, this is what will be presented. Adapting these proofs to the
discrete case is a fairly elementary exercise.

The key computation from \cite{dhks} is the following, whose proof is straightforward.

\begin{lemma}\label{L:variation}
If $H$ and $V$ are self-adjoint operators and $f$ and $g$ are vectors in the form domains
of both operators, then
\begin{align*}
& \langle f+\varepsilon g | H + V | f +\varepsilon  g \rangle +
    \langle f-\varepsilon g | H - V | f-\varepsilon g \rangle \\
{}={}& 2\langle f | H | f \rangle + 4\varepsilon \Re \langle f | V | g \rangle
    + 2\varepsilon^2 \langle g | H | g \rangle.
\end{align*}
\end{lemma}

We will also need the following little computational lemma that appears to go back to
Jacobi \cite{jac} (see also Courant-Hilbert \cite[p.~458]{ch}):

\begin{lemma}\label{L:ACalc}
Let $a$ be an $H^1=W^{1,2}$ function of compact support on $\R^d$. If $Q$ is locally
$L^1$ and $\psi\in W^{1,1}_{\mathrm{loc}}$ is a real-valued solution of $-\nabla^2\psi +
Q \psi = 0$ then
\begin{equation}\label{E:ACalc2}
  \int \|\nabla (a\psi)\|^2 + Q (a \psi)^2 = \int \|\nabla a\|^2 \psi^2 .
\end{equation}
\end{lemma}

\begin{proof}
Integrating by parts and using $-\nabla^2 \psi + Q \psi = 0$,
\begin{align*}
\int a^2 \|\nabla \psi\|^2 = - \int \psi \nabla \cdot (a^2 \nabla \psi)
    = - \int 2 a \psi (\nabla a) \cdot (\nabla \psi) + Q a^2 \psi^2
\end{align*}
and consequently
\begin{align*}
\int \|\nabla (a\psi)\|^2 + Q (a \psi)^2 & = \int \|a \nabla \psi + (\nabla a) \psi\|^2 + Q a^2 \psi^2 \\
&= \int a^2 \|\nabla \psi\|^2 + 2 a \psi (\nabla a) \cdot (\nabla \psi) + \|\nabla a\|^2 \psi^2 + Q a^2 \psi^2 \\
&=\int \|\nabla a\|^2 \psi^2
\end{align*}
as promised.
\end{proof}

\begin{proof}[Proof of Theorem~\ref{T:gscontgen2}.]
Without loss of generality, we assume that the ground state energy of $H_Q$ is zero. Let
$\psi$ be a bounded positive ground state for $H_Q$ and assume that $V \not\equiv 0$. Then
there exist $M>0$ and a smooth function $g$, supported in the ball of radius $M$ centered at the
origin, such that
\begin{equation}\label{E:pickg}
\int \psi V g < 0.
\end{equation}

Given $N > M$, define $a$ as follows: in one dimension,
\begin{equation}
a(x) = \begin{cases}  0 & |x| \geq N \\
    1 & |x| \leq M \\
    1-\frac{|x|-M}{N-M} & M < |x| < N
    \end{cases}
\end{equation}
and in two dimensions,
\begin{equation}
a(x) = \begin{cases}  0 & |x| \geq N \\
    1 & |x| \leq M \\
    \displaystyle \frac{\log N - \log |x|}{\log N - \log M} & M < |x| < N.
    \end{cases}
\end{equation}
Applying Lemma~\ref{L:ACalc} with $f=a\psi$, we obtain
\begin{align*}
\langle f | H | f \rangle = \int \|\nabla f\|^2 + Q f^2
= \int \|\nabla a\|^2 \psi^2 \lesssim \int \|\nabla a\|^2.
\end{align*}
In both one and two dimensions, the right-hand side converges to zero as $N\to\infty$;
in one dimension it is easily seen to be $O(N^{-1})$, for two dimensions,
\begin{align*}
\int \|\nabla a\|^2 &= \int \|\nabla a\|^2 \, r \, dr \, d\theta \\
&= 2 \pi [\log (N/M)]^{-2} \int_M^N r^{-1} dr \\
&= 2 \pi [\log (N/M)]^{-1}.
\end{align*}
This shows that $\langle f | H | f \rangle \to 0$ as $N \to \infty$.

By \eqref{E:pickg}, we can choose $\varepsilon > 0$ small enough so that
$$ 4\varepsilon \Re \langle f | V | g \rangle + 2\varepsilon^2 \langle g | H | g \rangle =
4 \varepsilon \Re \langle \psi | V | g \rangle + 2\varepsilon^2 \langle g | H | g \rangle < 0.
$$
Therefore, Lemma~\ref{L:variation} shows that, for $N$ large enough,
$$
\langle f+\varepsilon g | H_Q + V | f +\varepsilon  g \rangle + \langle f-\varepsilon g |
H_Q - V | f-\varepsilon g \rangle < 0.
$$
Thus, at least one of $H_{Q \pm V}$ has spectrum below zero.
\end{proof}

The proof shows that we need not have assumed that the ground state is bounded, but
merely has sufficiently slow growth at infinity. For example,
$|\psi(r,\theta)|=o(\log(r))$ would suffice in two dimensions.  In one dimension,
however, we can refine this idea and prove the following:

\begin{theorem}\label{T:gscontgen}
Suppose that $Q \in L^2_\mathrm{loc}(\R)$ and the operator $H_Q$ has a positive ground state
whose reciprocal is not square-integrable both at $+\infty$ and at $- \infty$. If $V \in
L^2_\mathrm{loc}(\R)$ and both $H_{Q \pm V}$ are bounded below by the ground state energy of
$H_Q$, then $V \equiv 0$.
\end{theorem}

\begin{proof}
The main idea is to choose $a$ in a manner that is adapted to $\psi$:
\begin{equation}
a(x) = \begin{cases}  0 & |x| \geq N \\
    1 & |x| \leq M \\
    \displaystyle 1-\frac{\int_M^x \psi(t)^{-2} \, dt}{\int_M^N \psi(t)^{-2} \, dt} & M < x < N \\[4mm]
    \displaystyle 1-\frac{\int_x^{-M} \psi(t)^{-2} \, dt}{\int_{-N}^{-M} \psi(t)^{-2} \, dt} & -N < x < -M.
    \end{cases}
\end{equation}
This necessitates only one change to the proof, namely,
the calculation which shows that
$\langle f | H | f \rangle \to 0$ as $N \to \infty$.

Note that $a$ is compactly supported and belongs to $H^1$; in fact, $a'\in L^\infty$ (as a distribution).
Therefore, we can apply Lemma~\ref{L:ACalc} to obtain
\begin{align*}
\int (f')^2 + Q f^2 \,dx &= \int (a')^2 \psi^2 \, dx \\
&= \int_M^N \frac{dx}{\psi(x)^2 \left\{\int_M^N \psi(t)^{-2} \, dt\right\}^2} +
    \int_{-N}^{-M} \frac{dx}{\psi(x)^2 \left\{\int_{-N}^{-M} \psi(t)^{-2} \, dt\right\}^2} \\
&= \biggl\{\int_M^N \psi(t)^{-2} \, dt\biggr\}^{-1} +
    \biggl\{\int_{-N}^{-M} \psi(t)^{-2} \, dt\biggr\}^{-1}.
\end{align*}
By assumption, the right-hand side goes to zero as $N \to \infty$.
\end{proof}

\end{document}